\documentclass[twocolumn,english]{article}

\usepackage{graphicx}

\makeatletter

\usepackage{ol2}
\usepackage[draft]{hyperref}
\usepackage{amsmath}

\makeatother

\begin{document}
\twocolumn[

\title{\textit{Shaping Laguerre-Gaussian laser modes with binary gratings
using a Digital Micromirror Device} }

\author{Vitaly Lerner, David Shwa, Yehonathan Drori, and Nadav Katz}

\maketitle
\address{Racah Institute of Physics, Hebrew University of Jerusalem}
\begin{abstract}
Laguerre-Gaussian (LG) beams are used in many research fields, such
as microscopy, laser cavity modes and optical tweezing. We develop
a holographic method of generating pure LG modes (amplitude and phase) with a binary amplitude-only Digital
Micromirror Device (DMD), as an alternative to the commonly used phase-only
Spatial Light Modulator. The advantages of such a DMD include
very high frame rates, low cost and high damage thresholds. We show
that the propagating shaped beams are self-similar and their phase
fronts are of helical shape as demanded. We estimate the purity of
the resultant beams to be above 94\%.
\end{abstract}
]

%\section*{Introduction}

Laguerre-Gaussian (LG) laser beam modes are solutions of the scalar Helmoltz
equation under the paraxial approximation \cite{AllenPadgett1995}.
Since LG beam modes were shown to carry an angular momentum \cite{angular_momentum}, they have been intensively investigated.
LG modes are widely used in many research
fields, including optical tweezing and atom guiding \cite{tweezers,atom-guiding},
second-harmonic generation \cite{Second_Harmonic}, quantum information
and communication \cite{quantum_info1}.

%****************HERE WORK*******************
Much of the work done on beam shaping has been carried
out recently with phase-only Spatial Light Modulators (SLMs)\cite{quant-info-3_SLM1,Mats-h-h,Kennedy}.
An alternative to this method is an amplitude control, using a Digital
Micromirror Device (DMD).
%C references changed
The most significant advantage
of the DMD over Liquid Crystal on Silicon (LCoS) SLM is frame rate \cite{Crystal,Dlp2011a}.
%CE
LG modes have been generated using static amplitude holograms\cite{Allen_Padgett_2008},
while here we focus on programmable patterns. LG amplitude \cite{Liyang2012} and helical phase \cite{Yu-Xuan} have been generated separately using a DMD, while we focus on pure LG modes (both amplitude and phase).

Mathematically, LG modes are described in  cylindrical coordinates, where we rescale $r,\, z$
by the beam width $w_{0}$ and corresponding Rayleigh range $z_{0}=\frac{\pi w_{0}^{2}}{\lambda}$
: $\rho=\frac{r}{w_{0}};\:\zeta=\frac{z}{z_{0}}$. The width of all such
Gaussian beams, at distance $\zeta$ from
the waist is $R(\zeta)=\sqrt{1+4\zeta^{2}}$ . We can write the electric
field of a LG mode \cite{AllenPadgett1995} using generalized Laguerre polynomials $L_p^l(x)$, separated to amplitude and phase, omitting the global normalization
constants and the global phase due to the propagation distance :

\begin{equation}
\begin{array}{r}
\left|u_{p}^{l}\left(\rho,\varphi,\zeta\right)\right|=\left(\frac{2\rho^{2}}{R^{2}(\zeta)}\right)^{|l|}
L_{p}^{|l|}\left(\frac{2\rho^{2}}{R^{2}(\zeta)}\right)
%\textnormal{e}^{-\frac{\rho^{2}}{R^{2}(\zeta)}}
\exp\left(-\frac{\rho^{2}}{R^{2}(\zeta)}\right)
\end{array}\label{eq:LG_INTE}
\end{equation}

\begin{equation}
\Phi\left\{ u_{p}^{l}\left(\rho,\varphi,\zeta\right)\right\} =l\varphi+\frac{\rho^{2}}{R^{2}(\zeta)}\zeta\label{eq:LG_PHS}
\end{equation}

%C text omitted
%It is easily seen from Eq. (\ref{eq:LG_INTE}) that the intensity profile is conserved with distance, up to the beam width which grows due to diffraction.
%CE

%\paragraph*{Fork-like patterns}

\begin{figure}[h]
\includegraphics[width=3.25in]{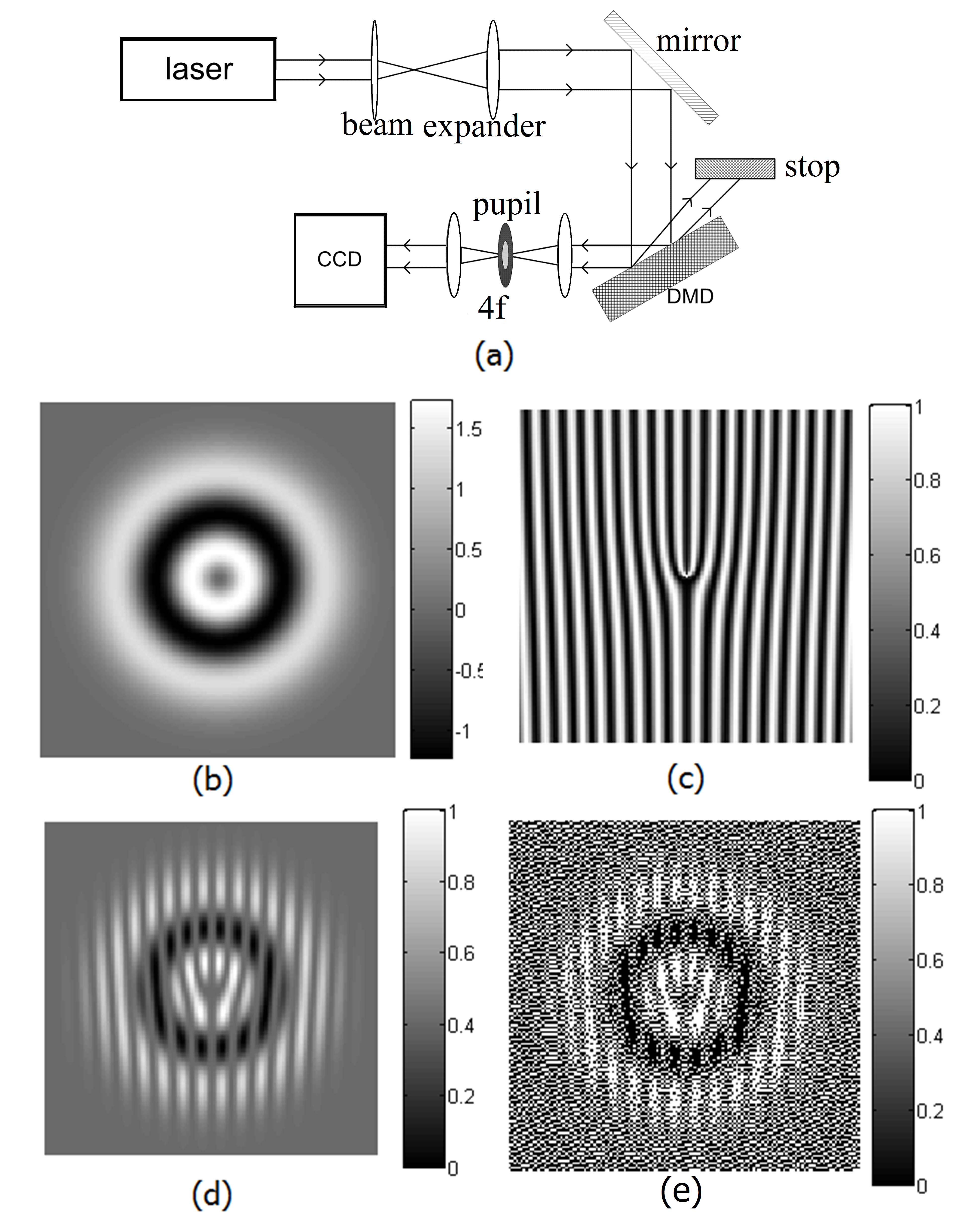}

\noindent \caption{(a) Schematic of the experimental setup. (b)-(e) Shaping amplitude and phase of $LG_{2}^{2}$ mode (all patterns are
240x240 px, corresponding to 1.8x1.8 mm): (b)ideal amplitude; (c)fork-like
hologram for l=2; (d) normalized multiplication of (b) by (c); (e)dithered
pattern loaded on the DMD. Note that the background corresponds to
half filling to allow for negative amplitude of alternate rings in the
LG pattern.
\label{fig:Schematic-of-the}}
\end{figure}

While shaping a phase front (Eq. (\ref{eq:LG_PHS}) ) is a relatively
simple procedure when using a phase-only SLM, e.g. LCoS, shaping an
amplitude pattern is a matter of considerable algorithmic work\cite{Mats-h-h}.
In our case, the situation is opposite. We use an amplitude-only SLM.
It is very easy to load a pattern on the DMD and to observe the same
pattern at the imaging plane. However, a phase front cannot be created
directly. A convenient technique \cite{Yu-Xuan,Molina-Terriza2007,Bekshaev2010,Allen_Padgett_2008}
to create light vortices of type $\Phi_{l}(\varphi)=l\varphi$ is to use
a fork-like pattern, e.g. Fig. \ref{fig:Schematic-of-the}(c).
The pattern is a holographic interference of an LG beam with a
unitary planar beam. Assuming that the angle between the beams is
$\alpha$ and the LG beam is at its waist ($\zeta=0,\: R(\zeta=0)=1$),
the phase front becomes

\begin{equation}
\Phi_{l}(\rho,\varphi)=l\varphi+\frac{2\pi}{\lambda}\rho\cos\varphi\sin\alpha\label{eq:fork_phase}
\end{equation}

The diffraction grating we use for creating such a front is

\begin{equation}
I_{l}(\rho,\varphi)=\cos\left(\Phi_{l}(\rho,\varphi)\right)\label{eq:fork_smooth}
\end{equation}

%\section*{Experimental Setup}

%\paragraph{Digital micromirror device}
We use a commercial DMD device, adapted and mounted for optical experiments \cite{Methods}
 The DMD consists of 480x320 micromirrors, each 7.6x7.6
microns in size. Each mirror corresponds to a certain pixel and it
is held in either of two angular positions: $+12^o$ (on, or {}``white''
state) and $-12^o$ (off, or {}``black'') state\cite{DLPTM}. The
DMD serves as a programmable spatial filter since the light reflected
from and diffracted by the mirrors corresponding to the black pixels
is filtered out (Fig. \ref{fig:Schematic-of-the}).

%\paragraph{Efficiency}

The maximal efficiency of a grey amplitude hologram is known to be $\frac{1}{16}=6.25\%$, while
for binary amplitude it can reach $\frac{1}{\pi^2}\sim 10.1\%$  \cite{LohmannBrown1969,Lohmann1967}. Due to the complex three dimensional
structure of the DMD surface, the pattern effectively consists of
two gratings: the micromirrors grating-like structure and the information
grating.When all pixels are on, at a particular angle, we concentrate more than 88\% \cite{DLPTM,DMD} of the diffracted
light into a single order.We measure an additional loss of about 40\% of the total power of the incident
beam. After applying the information grating between 1\% and 5\% of the incident
beam intensity is concentrated in an output beam.

%\paragraph*{Optical Setup}

In order to obtain a desired pattern at the imaging plane, the first
order beam is transferred and other diffraction beams are filtered
out by the pupil (Fig. \ref{fig:Schematic-of-the}(a)). In our experiments 40cm away
from the DMD the beams are well separated. However, in order to reproduce
the near field, we use the imaging system.

%\paragraph*{Phase modulation and simple vortices}

Two optical vortices, i.e. diffractions of $\pm$1\textsuperscript{st}
orders can be are created by applying a fork-like pattern (Eq. (\ref{eq:fork_smooth}),
Fig. \ref{fig:Schematic-of-the}(c) ). A similar method
is described in Ref. \cite{Yu-Xuan}, where slow grey scale pixel
modulation was used. However, in many applications stable vortex patterns
are required. Therefore we avoid modulating pixels in time, and resort
to pixel dithering \cite{dither1}. The dithering leads to effective
grey scale amplitude control.

%\paragraph*{Amplitude modulation and compensation due to an incident beam}

The fork-like patterns are used to create vortices with no direct
control of the amplitude\textcolor{black}{. Ring-like amplitude has
been achieved in previous works, e.g. \cite{Allen_Padgett_2008},
by $\frac{\pi}{2}$-shifts at the zeros of LG radial component of
amplitude distribution. We apply a different method, achieving significantly higher purity of the LG modes. We separate the beam shape to its amplitude and phase components (Eq.
(\ref{eq:LG_INTE}) and (\ref{eq:fork_phase})). We multiply the fork-like
pattern by the amplitude distribution and dither the resultant pattern
(Fig. \ref{fig:Schematic-of-the}). }

Previous works \cite{Allen_Padgett_2008,Mats-h-h}
have shown that the purity of the LG modes was significantly affected
by the ratio of the normalization radius of the hologram and the input
Gaussian beam. Even at the optimal value of this ratio, the
purity of the LG modes with $p>1$ was low when amplitude-only hologram
was applied to Gaussian beam \cite{Allen_Padgett_2008}. We optimize the purity
of the output beams by changing the applied pattern:

\begin{equation}
\begin{array}{r}
\left|\overset{corr}{u_{p}^{l}}\left(r,\varphi,z=0\right)\right|=
\left(\frac{2r^{2}}{w^{2}}\right)^{l} L_{p}^{l}\left(\frac{2r^{2}}{w^{2}}\right)
%\textnormal{e}^{-p_{0}\cdot\frac{r^{2}}{w^{2}}}
\exp\left(-p_{0}\cdot\frac{r^{2}}{w^{2}}\right)
\end{array}\label{eq:final_amp_pattern}
\end{equation}
, where  $p_0=\frac{w^2}{w_{env}^2}=1-\frac{w^2}{w_i^2}$
is a correction parameter, $w_i$ is the width of the incident Gaussian beam, $w_{env}$ is the corrected width of the Gaussian envelope of the pattern and $w$ is the desired width of the shaped LG mode.
The correction parameter can be calculated from measurements and substituted back
to Eq. (\ref{eq:final_amp_pattern}), but it should be fine-tuned
empirically.

\textcolor{black}{We measure a minimal radius at which there is no
distortion of the resultant mode and a good off/on ratio to be 12
pixels (corresponding to about 90$\mu m$).}

\begin{figure}[h]
\includegraphics[width=3.25in]{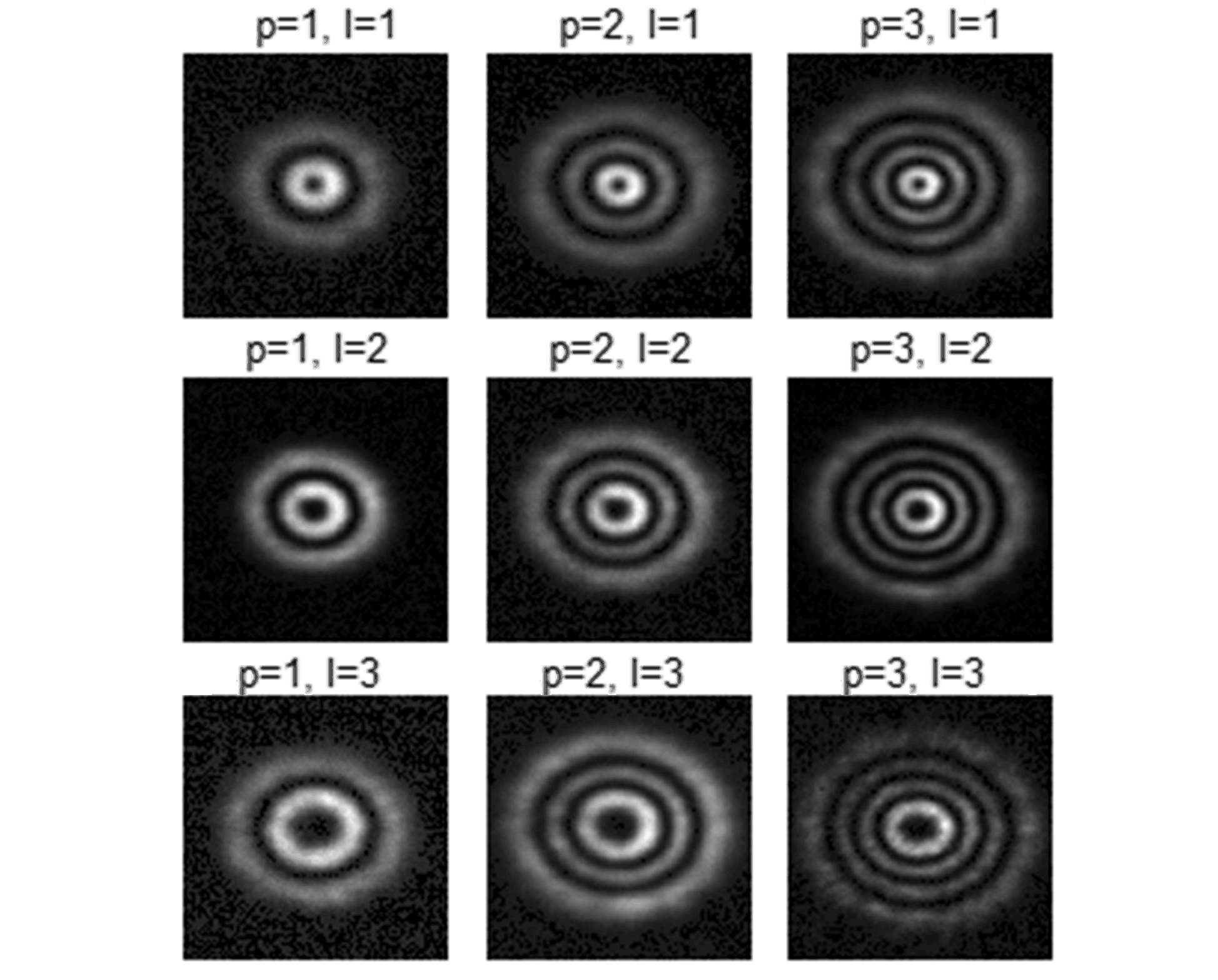}
\caption{Intensities in far field. Radius of the outer ring of $LG_{3}^{3}$
beam is 700$\mu m$\label{fig:Intensities-far-field}}
\end{figure}

\begin{figure}[h]
\includegraphics[width=3.25in]{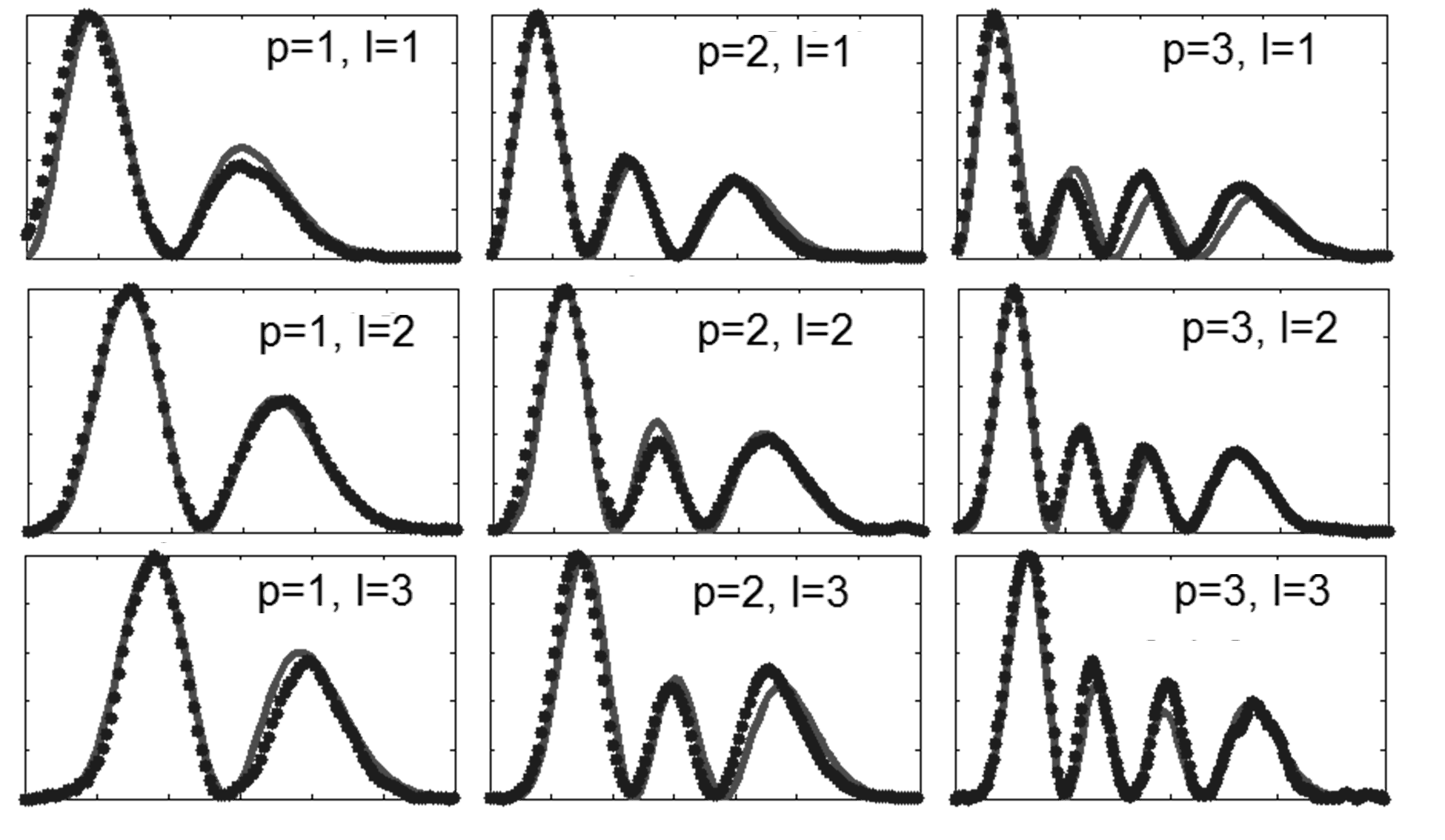}
\caption{Radial profiles of intensity obtained from cross sections of Fig. \ref{fig:Intensities-far-field}(dotted)
vs ideal LG intensity profiles (solid)\label{fig:profiles@far-field}}
\end{figure}

\begin{figure}[h]
\includegraphics[width=3.25in]{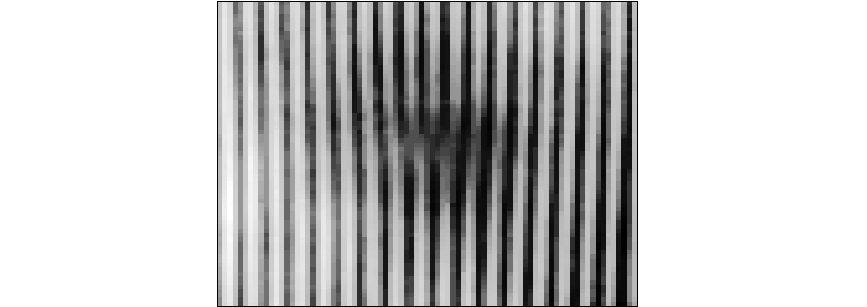}

\caption{: Interference of an $LG_{0}^{1}$ beam with an off-axis Gaussian beam in far field. Physical height: 100$\mu m$\label{fig:phase_amp_all} }
\end{figure}

The results obtained at the imaging plane and in the near field show no significant
difference from the far field, proving that the obtained
modes are true LG modes (Eq. (\ref{eq:LG_INTE})) We present (Fig. \ref{fig:Intensities-far-field}) the results
in far field \cite{elipticity}.
%C text omitted
In order to ensure that the far field is measured, and to focus the beam on the camera, we place a convex lens (f=20 cm) exactly 20 cm from the camera and 50 cm from the DMD. The Rayleigh range of the resultant beam was 12 cm.
%CE
In order to study intensity profiles, we plotted cross-sections along with the ideal
modes intensity profiles (Fig. \ref{fig:profiles@far-field}).
In order to confirm the helical phase front, the resultant beams
are interfered with an off-axis Gaussian beam. The center of the
resultant pattern is shown in Fig. \ref{fig:phase_amp_all}, and
its fork-like shape confirms \cite{Molina-Terriza2007} that in the far-field the beam has a helical phase front.

The experimental results fit almost perfectly the ideal intensity
profiles. The off:on ratio (intensity at the origin divided by maximal intensity) varies between 1:100
and 1:1000.

In order to give a quantitative measure of the beams'
quality, we calculate mode purity in a manner similar to that of Ref. \cite{Kennedy}. Since we do not measure directly either amplitude or phase (only intensity), we use
a simulation of far-field Fraunhoffer diffraction to estimate the
mode purity. As an input we use the same pattern loaded on the DMD
multiplied by a Gaussian beam. Both simulation
field $u_{p,sim}^{l}(r,\varphi)$ and ideal one $u_{p,id}^{l}(r,\varphi)$
are normalized. We define purity as

\begin{equation}
P_{p}^{l}=\left| \iint{u_{p,sim}^{l}}^{*}  \cdot u_{p,id}^{l} \cdot rdrd\varphi \right|^2
\end{equation}

The results for different modes vary from 95\% ($P_{3}^{3}$) to 97\%
($P_{2}^{1}$).

A simpler direct correlation overlap of the experimental results (intensities) with the
ideal intensities (see Fig. \ref{fig:profiles@far-field}) yields slightly lower results, 94\% to 96\%. This indicates
that experimentally we are very close to the optimal reachable result
using this method.

%\section*{Discussion and conclusions}

In summary, we use a DMD as an amplitude-only SLM and demonstrate the shaping
of Laguerre Gaussian modes. We achieve high purity, above 94\%, of the modes, and we show that such purity is nearly optimal
when applying our method. Further improvements may include employing
a higher resolution DMD, and a fine adjustment of patterns in closed
loop control.

%\section*{Aknowledgments}

We thank Nir Davidson and Hagai Eisenberg for helpful suggestions and discussion. This research was supported by ISF grant 1248/10.

%************************************
%************************************
%**********FOURTH PAGE****************
%****************************************
%\newpage
\bibliographystyle{osajnl}
%\bibliography{LGBeamsLONG}
\bibliographystyle{ol}
%\bibliography{LGBeams}

\begin{thebibliography}{10}
\bibitem{elipticity}
{The barely observable elipticity is due to small imperfactions in our tilt
  angle adjustments, and input Gaussian beam. However, no splitting in l=2 and
  higher modes is observed indicating that this imperfection is negligable.}
\end{thebibliography}

\end{document}